\documentstyle[prl,aps,psfig]{revtex}
\tighten
\begin{document}
\normalsize

\title{Ratio of Hadronic Decay Rates of $J/\psi$ and $\psi(2S)$ and 
 the $\rho\pi$ Puzzle}
\author{Y.~F.~Gu$^{(1),\dag}$, X.~H.~Li$^{(2)}$}

\address{
(1)Institute of High Energy Physics,
  Beijing 100039, People's Republic of China \\
(2)Physics Department, University of California, Riverside, CA 92521, USA}
\date{\today}

\maketitle

\begin{abstract}
 The so-called $\rho\pi$ puzzle of $J/\psi$ and $\psi(2S)$ decays is
 examined using the experimental data available to date.  Two different
 approaches were taken to estimate the ratio of $J/\psi$ and $\psi(2S)$
 hadronic decay rates.  While one of the estimates could not yield the
 exact ratio of $\psi(2S)$ to $J/\psi$ inclusive hadronic decay rates, 
 the other, based on a computation of the inclusive ggg decay rate for 
 $\psi(2S)$ ($J/\psi$) by subtracting other decay rates from the total 
 decay rate, differs by two standard deviations from the naive prediction 
 of perturbative QCD, even though its central value is nearly twice as 
 large as what was naively expected.  A comparison between this ratio, 
 upon making corrections for specific exclusive two-body decay modes, 
 and the corresponding experimental data confirms the puzzles in 
 $J/\psi$ and $\psi(2S)$ decays.  We find from our analysis that the 
 exclusively reconstructed hadronic decays of the $\psi(2S)$ account for 
 only a small fraction of its total decays, and a ratio exceeding the 
 above estimate should be expected to occur for a considerable number 
 of the remaining decay channels.  We also show that the recent new results 
 from the BES experiment provide crucial tests of various theoretical 
 models proposed to explain the puzzle.   
 \end{abstract}

 \vspace{.6cm}

 One of the outstanding problems in heavy quarkonium physics 
 is the strong suppression of the $\psi(2S)$ decays to vector plus pseudoscalar-meson (VP)
 final states, $\rho\pi$ and $K^+\overline{K}^{\ast -}+c.c.$,
 which is referred to as the $\rho\pi$ puzzle~\cite{puz}.
 Following the first observation of this anomaly~\cite{mark}, meagre 
 experimental progress was made over the years, and theoretical
 analysis based on limited data often lead to unsatisfactory, sometimes
 premature, inferences.  The situation has been changed dramatically in the
 last few years.  A wealth of interesting new information, which extended 
 the puzzle considerably, has emerged from intense studies of $\psi(2S)$ 
 hadronic decays at the BES experiment, using a large sample of 3.79 
 million $\psi(2S)$ decays~\cite{harris}.  It is hoped for that new concerted  
 efforts on both the theoretical and experimental side would eventually lead 
 to a solution of this long-standing conundrum.

 In this paper we seek to examine the $\rho\pi$ puzzle based purely on
 existing experimental data.  We begin with an analysis for estimating the
 ratio of hadronic decay rates of $J/\psi$ and $\psi(2S)$, which we shall
 denote by $Q$, by using the data compiled by the Particle Data Group~\cite{pdg}, 
 in an attempt to avoid as many theoretical ambiguities as possible in the analysis.  
 Two different approaches to this estimate are performed. First we compare the 
 results between themselves, and then the naive prediction of perturbative QCD (PQCD)
 is used. Subsequently, possible corrections to $Q$ are discussed, as they
 associate with specific exclusive decay modes, and the corrected values of $Q$ 
 are used as standards to compare with the corresponding experimental data.  
 Comments on the issue of $J/\psi$ and $\psi(2S)$ decays to multi-hadron final states and on the 
 potential similarity of the $\eta_{c}$-$\eta_{c}(2S)$ decays to the $\rho\pi$ 
 puzzle are profusely added. Finally, various theoretical models are discussed in order to
 offer a greater understanding of the $\rho\pi$ puzzle, in light of recent BES results.
  
    Conventionally, measured ratios of $\psi(2S)$ to $J/\psi$ branching
 fractions for specific exclusive hadronic decays are compared with 
 the naive prediction of PQCD, the so-called ``15\% rule''.
 In the framework of PQCD~\cite{pqcd}, one expects $J/\psi$($\psi(2S)$) 
 to decay to hadrons via three gluons, or a single direct photon.  
 In either case, the partial width of the decay is proportional 
 to $|\Psi(0)|^{2}$, where $|\Psi(0)|$ is the wave function at the origin 
 in the nonrelativistic quark model of the $c\overline{c}$ quark state.  
 Thus one finds that 
\begin{equation}
Q_h\equiv \frac{B(\psi(2S)\rightarrow ggg)}{B(J/\psi \rightarrow ggg)}=
\frac{\alpha_{s}^{3}(\psi(2S))}{\alpha_{s}^{3}(J/\psi)}      
\frac{B(\psi(2S)\rightarrow e^{+}e^{-})}{B(J/\psi\rightarrow e^{+}e^{-})}=
(14.8\pm2.2)\%,
\end{equation}
 where the new world averages of the leptonic branching fractions 
 are used (see Table 1). This is assuming that 
 the strong coupling constants are equivalent, i.e., $\alpha_{s}(\psi(2S))$=$\alpha_{s}(J/\psi)$.  
 Taking the running constant $\alpha_{s}$ into
 account~\cite{alphas}, the ratio $Q_h$ becomes
\begin{equation}
Q_h=(12.5\pm1.9)\%.
\end{equation}
 The MARKII experiment first compared the theoretical prediction of this value
 ($(12.2\pm2.4)\%$ then used) with measurements for a number of exclusive hadronic 
 decays of the $J/\psi$ and the $\psi(2S)$, thus revealing the $\rho\pi$ puzzle~\cite{mark}.
 
 However, this naive prediction suffers several apparent approximations. 
 Higher order corrections, which may not even be small, are not included in this 
 calculation. For example, a first order correction  
 to the branching fraction of $J/\psi\rightarrow e^{+}e^{-}$ could be 
 50$\%$ of the lowest term if one were to use $\alpha_{s}(m_{J/\psi}) 
 \sim 0.2$~\cite{kwong}. 
 The relativistic effect is also ignored.  Since the mass difference between 
 $J/\psi$ and $\psi(2S)$ is around 
 $20\%$ and $<v^{2}/c^{2}>\sim0.24$ for $J/\psi$, this correction
 may be at the same level as the lowest order~\cite{kwong}. 
 The inclusion of the finite size of the decay vertex 
 will significantly reduce the ggg decay width of $J/\psi$~\cite{hcc}.   
 Moreover, the effect of non-perturbative dynamics is neglected, 
 the size of which is hard to estimate. Therefore, people may question the 
 validity of the ``15$\%$ rule'' as a serious benchmark for 
 comparing experimental data. 

 We present here two approaches to estimate the ratio $Q_h$, using the 
 data as displayed in Table I-III.  The first approach is based on an
 assumption that the decays of the $J/\psi$ and $\psi(2S)$ 
 in the lowest order of QCD are classified into  
 hadronic decays (ggg),
 electromagnetic decays ($\gamma^{\ast}$), 
 radiative decays into light hadrons ($\gamma gg$), and decays to lower 
 mass charmonium states(c$\overline{c}X$)~\cite{lknw,yfg}.  
 Thus, using the relation
 $ B(ggg)+B(\gamma gg)+B(\gamma^{\ast})+B(c\overline{c}X)=1$,
 one can derive B(ggg)+B($\gamma gg$) by subtracting B($\gamma^{\ast}$) and 
 B(c$\overline{c}X$) from unity. 

 The electromagnetic decay channels of the $J/\psi$ produce hadrons, $e^{+}e^{-}$ 
 and $\mu^{+}\mu^{-}$ as final states. Besides these channels, the
 electromagnetic decays of $\psi(2S)$ also include the $\tau^{+}\tau^{-}$
 as a final state.  
 The experimental data is summarized  
 in Table 1, where the branching fraction of $\psi(2S)\rightarrow \tau^{+}\tau^{-}$
 is a recent measurement of the BES experiment~\cite{tau}, whereas the other data 
 was taken from the Particle Data Group~\cite{pdg,com}.  
 The total contributions to the electromagnetic decays 
 of $J/\psi$ and $\psi(2S)$ are then given as
 $B(J/\psi \rightarrow \gamma^{\ast})=(28.81\pm2.00)\%$ and 
 $B(\psi(2S) \rightarrow \gamma^{\ast})=(5.08\pm0.55)\%$, respectively. 

 As regards to the decay into lower mass charmonium states, the $J/\psi$ 
 has only one radiative decay channel into $\eta_{c}$, whereas the $\psi(2S)$ can 
 decay into a number of other final states, i.e. 
    $\pi^{+}\pi^{-}J/\psi$,
    $\pi^{0}\pi^{0}J/\psi$,
    $\pi^{0}J/\psi$,
    $\eta J/\psi$,
    $\gamma\chi_{c0}$,
    $\gamma\chi_{c1}$,
    $\gamma\chi_{c2}$,
    $\gamma\eta_{c}$,
    $\gamma\eta_{c}(2S)$ and 
    $1^{1}P_{1}+ X$.   
 The decay rates of the last two channels are faint 
 and  thus are neglected in our calculation. 
 The experimental data summarized in Table 2 
 are all taken from PDG~\cite{pdg}.
 Using this data we calculated the total 
 contribution to $c\overline{c}X$:  $B(J/\psi \rightarrow c\overline{c}X)=(1.3 \pm0.4)\%$ and 
 $B(\psi(2S)\rightarrow c\overline{c}X)=(78.1\pm3.9)\%$,  
 respectively.
 
 By deducting the contributions B($\gamma^{\ast}$) and B(c$\overline{c}X$), 
 we find that  
 $B(J/\psi \rightarrow ggg)+B(J/\psi \rightarrow \gamma gg)=(69.9\pm2.0)\%$ 
 and $B(\psi(2S) \rightarrow ggg)+B(\psi(2S) \rightarrow \gamma gg)=
 (16.8\pm3.9)\%$.  Therefore the ratio of branching fractions 
 of $\psi(2S)$ to $J/\psi$ decays into hadrons is given by
\begin{equation}
Q_1= \frac{B(\psi(2S) \rightarrow ggg)+B(\psi(2S) \rightarrow \gamma gg)}
    {B(J/\psi \rightarrow ggg)+B(J/\psi \rightarrow \gamma gg)}
     = (24.0\pm5.6)\%.
\end{equation}
 The relation between the decay rates of ggg and $\gamma gg$ is readily 
 calculated in PQCD to the first order as~\cite{kwong}
\begin{equation}
\frac{\Gamma(J/\psi\rightarrow\gamma gg)}{\Gamma(J/\psi\rightarrow ggg)}=
\frac{16}{5}\frac{\alpha}{\alpha_{s}(m_{c})}(1-2.9\frac{\alpha_{s}}{\pi}). 
\end{equation}
 Using $\alpha_{s}(m_{c})=0.28$, one can estimate   
 $\Gamma(J/\psi\rightarrow\gamma gg)/\Gamma(J/\psi\rightarrow ggg)
 \simeq 0.062$.  A similar relation can be deduced for the $\psi(2S)$ decays. 
 Thus one expects that "24\% ratio" stands well for either ggg mode or 
 $\gamma gg$ mode. 
   
 The other approach is to use the data on branching fractions for hadronic 
 decays in final states containing pions, kaons, and protons that have 
 already been measured for both $J/\psi$ and $\psi(2S)$.  They are 
 $3(\pi^{+}\pi^{-})\pi^{0}$,
 $2(\pi^{+}\pi^{-})\pi^{0}$,
 $\pi^{+}\pi^{-}\pi^{0}$,
 $\pi^{+}\pi^{-}K^{+}K^{-}$,
 $\pi^{+}\pi^{-}P\overline{P}$,
 $P\overline{P}$,
 $P\overline{P}\pi^{0}$ and $K^+K^-$. 
 Using the PDG data compiled in Table 3, we have 
 $\sum\limits_{i=1}^{8} B_{i}(J/\psi\rightarrow f_{i})=(9.43\pm0.72)\%$ and 
 $\sum\limits_{i=1}^{8} B_{i}(\psi(2S)\rightarrow f_{i})=(0.941\pm0.185)\%$. 
 It follows that
\begin{equation}
Q_2=\sum\limits_{i=1}^{8} B_{i}(\psi(2S)\rightarrow f_{i})/
\sum\limits_{i=1}^{8} B_{i}(J/\psi \rightarrow f_{i})=(10.0\pm2.1)\%.
\end{equation}

 We note that the results obtained by the two approaches vary
 considerably.  However, a comparison of the values of
 the total branching fraction for
 $\psi(2S) \rightarrow ggg$ computed by the
 two approaches indicates that only a small fraction ($\sim 6\%$) of the
 exclusive hadronic decays of $\psi(2S)$
 have been reconstructed experimentally.  It is thus obvious that $Q_2$ is
 not the exact ratio of $\psi(2S)$ to $J/\psi$ inclusive hadronic decay
 rates, but represents on average the ratio of the exclusive decay channels,
 as measured to date.  We therefore do not consider $Q_2$ any further.
 Nevertheless, the question persists as to where the remaining hadronic
 $\psi(2S)$ decay modes are and how the corresponding pattern of decays
 for $\psi(2S)$ to $J/\psi$ behaves.  It would be an intriguing experiment 
 task to search for those remaining channels that are in such final states as 
 those with higher multiplicities, or those with multi-neutral particles, 
 or even for remaining channels in non-$q\overline{q}$ states.
 In comparison with the naive PQCD expectations, the central value of $Q_1$
 is about a factor of two higher than that of $Q_{th}$, as stated in Eq. (2).
 However, the difference lies within the 2$\sigma$ error of $Q_1$ and is
 only marginally significant.  The substantial error of $Q_1$ is essentially
 due to the propagation of errors during the subtraction of the decay rate
 $B(\psi(2S)\rightarrow c\overline{c}X)$ from the total rate, although the
 total error of $B(\psi(2S)\rightarrow c\overline{c}X)$ itself merely 
 amounts to about 5$\%$.  Taking into account the apparent approximations
 to the naive expectations of PQCD as well, it seems to us premature to 
 regard this as a remarkable discrepancy that deserves serious 
 considerations.
   
 To use the ratio $Q_1$ for comparison with the experimental results of
 $\psi(2S)$ and $J/\psi$ decays, it should be noted that the estimate of
 $Q_1$, like the prediction of $Q_{th}$, is made for the total width for
 ggg decay, not for the partial widths of exclusive final states.
 Consequently, a number of
 corrections may be associated with specific exclusive decay modes:
 (I) It is shown that the $J/\psi$ and $\psi(2S)$ decays to
 $\omega f_2$, $\rho a_2$, $K^{\ast 0} \overline{K}_2^{\ast 0}+c.c.$ and
 $\phi f_2^{\prime} (1525)$ are hadron helicity conservation (HHC)
 allowed~\cite{gutuan},
 while that to $\rho\pi$ and $K^{\ast}\overline{K}$
 are HHC forbidden~\cite{sjb}.
 The general validity of the HHC at the charmonium mass scale is still
 an open question. 
 It is suggested that a critical test would be to measure
 the angular distributions of exclusive final states~\cite{sjb}.
 Existing measurements on angular
 distributions for $J/\psi$ decays into $p\overline{p}$,
 $\Lambda\overline{\Lambda}$, $\Sigma^{0}\overline
 {\Sigma}^{0}$ and $K^{0}_{S}K^{0}_{L}$~\cite{ang}
 are consistent with the HHC predictions (baryon pairs
 within $1-1.5$ standard deviations); however, no data is available for
 $\psi(2S)$ decays so far.
 Exclusive processes that violate helicity conservation are suppressed
 by powers of $m^{2}/s$ in QCD~\cite{sjb}.
 This would contribute a suppression factor
 $M^{2}_{J/\psi}/M^{2}_{\psi(2S)}=0.71$
 to the ratio of the $\psi(2S)$ to $J/\psi$ decay rates for
 final states $\rho\pi$ and $K^{\ast}\overline{K}$.
 On the other hand,  the  $\gamma\eta$ and $\gamma\eta'$ modes
 are allowed by the helicity selection rule, since
 helicity conservation applies only to the hadrons~\cite{gutuan}.
 (II) Exclusive reactions which involve hadrons with quarks
 or gluons in higher orbital angular momentum states  are suppressed by
 powers of $1/s=E^{-2}_{cm}$~\cite{sjb}.  This contributes a suppression factor
 of $M^2_{J/\psi}/M^2_{\psi(2S)}$ to the ratio of
 $\psi(2S)$ vs. $J/\psi$ branching fractions for decays into such final states 
 as
 $\omega f_2$, $\rho a_2$, $K^{\ast 0} \overline{K}_2^{\ast 0}+c.c.$ and
 $\phi f_2^{\prime} (1525)$,
 where a meson in P-wave state is included.
 (III) Ref.~\cite{sjb} reported another suppression
 arising from the asymptotic form factor
 which would be   $M^8_{J/\psi}/M^8_{\psi(2S)}=0.25$ for decays to
 $p\overline{p}$ channel.
 Contrary to these calculations,
 Ref.~\cite{bolz} evaluated the three-gluon contribution
 with the c-quark mass instead of the charmonium mass.
 As a consequence, the ratio of the $J/\psi$ and $\psi(2S)$ decay widths 
 is not scaled to the 8th power of the ratio of their masses in our calculations.

 Table 4 lists corrections to the ratio $Q_1$ for several exclusive
 hadronic decay channels.
 Experimental data from PDG~\cite{pdg}
 are also included for comparison.
 As seen from the table, the predicted corrected value of the ratio $Q_1$ for
 $b_1\pi$ is consistent with the experimental data, whereas the experimental
 ratio of $K_1^{\pm}(1270)K^{\mp}$ is enhanced as compared with the
 predicted value. The deviations of the measured ratios for decays into VP, 
 VT and other final states (even
 $p\overline{p}$) from the corrected ratios demonstrate suppressions, in this case.
 Note that the combination of all the above correction results lead to a
 substantial reduction of the ratio of $\psi(2S)$ to $J/\psi$ decay rates,
 well below 24\% for many of the exclusive hadronic decay channels (this is
 also compatible with the observation mentioned above that the value of
 $Q_2$ is much lower than that of $Q_1$).
 One should therefore conclude that a considerable number of other decay channels
 ought to have an enhancement with a ratio above 24\%, in order to make up for
 all these  suppressed channels.
 It is puzzling that so far there has only been one channel for the $\psi(2S)$ decays
 observed, the $K_1^{\pm}(1270)K^{\mp}$ channel, which is enhanced relative to
 the $J/\psi$~\cite{besap}. Further systematic 
 study of $\psi(2S)$ decays are anxiously awaited.

 As is seen from the above analysis, we have restricted our comparison only to
 decays to two-body final states.  What if one makes a comparison for
 decays leading to three or more hadrons?  MARKII did make such a
 comparison in their original work~\cite{mark}
 and claimed that the ratio $Q$ for decay modes such as
 $p\overline{p}\pi^{0}$, $p\overline{p}\pi^{+}\pi^{-}$,
 $K^{+}K^{-}\pi^{+}\pi^{-}$, 2$(\pi^{+}\pi^{-})\pi^{0}$,
 and 3$(\pi^{+}\pi^{-})\pi^{0}$,
 is consistent with the naive theoretical expectations.
 However, it should be pointed out that most of these multi-hadron final
 states in fact include sums of several two-body intermediate states.
 One thus observes a mixed effect which may not deviate noticeably
 from the expected value of $Q$, even if a few of the two-body intermediate
 states are severely suppressed.  For example, the decay
 $J/\psi\rightarrow 2(\pi^{+}\pi^{-})\pi^{0}$ proceeds predominantly
 through intermediate states $b_1\pi$, $\omega f_2$, and $a_2\rho$.
 The observed $Q$ for this decay, as reported by MARKII~\cite{mark},
 was $(9.5\pm2.7)\%$, does deviate, though not quite significantly,
 from the ``$15\%$ rule''.  This results from the fact that two of these
 three two-body intermediate states are found to be anomalously 
 suppressed~\cite{besvt}.  Therefore, one must be always cautious about
 drawing conclusions from the comparison of decays 
 of multi-hadron final states. 

 An experimental situation similar to the $\rho\pi$ puzzle occurs in the
 decays of the $\eta_c$ in three vector meson (VV) cases, 
 such as $\rho\rho$, $K^{\ast}\overline K^{\ast}$ and $\phi\phi$,
 and in $p\overline{p}$.  These decays are all first-order forbidden
 by HHC in PQCD~\cite{sjb,anselm};  however, they are actually
 observed to occur with relatively large branching fractions~\cite{pdg}.
 It is thus interesting to look for the analogous decays
 of the $\eta_{c}(2S)$ and compare the ratio of the $\eta_{c}(2S)$ to
 $\eta_c$ branching fractions with the relation
 $B(\eta_{c}(2S)\rightarrow h)\simeq B(\eta_{c}\rightarrow h)$
 predicted by Chao et al~\cite{cgt}.
 In testing for helicity conservation, these decay modes
 for $\eta_c$ and its spin-singlet partner $\eta_{c}(2S)$ play the
 same role as the decay modes $\rho\pi$ and $K^{\ast}\overline{K}$ do in
 the case of $J/\psi$ and its spin-triplet partner $\psi(2S)$.
 The search for the $\eta_{c}(2S)$ is thus important
 not only because the $\eta_{c}(2S)$ is one of the two remaining states 
 of the charmonium family awaiting confirmation (or discovery) but also 
 because the study of its hadronic decays could shed light on the puzzle 
 of $J/\psi$ and $\psi(2S)$ decays. 

 We now move on to discuss various theoretical models made to explain the
 $\rho\pi$ puzzle as it is presently formulated.  Instead of a critical examination 
 of many theoretical arguments (which one can find in the 
 literature~\cite{puz,chao,fk,suz}), we will concentrate exclusively on 
 comparing the experimental results, mostly from the BES experiment, to 
 the predictions of these models, in an attempt to differentiate 
 between them. 

 The first explanation for the MARKII observation, as proposed by Hou and
 Soni~\cite{hs} and generalized later by Brodsky et al.~\cite{blt},
 is the postulate that the decay $J/\psi$, which violates the helicity
 selection rule of PQCD, is enhanced by the mixing of the
 $J/\psi$ with a vector glueball $O$ that decays preferentially
 to $\rho\pi$ and other VP channels.  The vector glueball $O$ is required
 to be fairly narrow and nearly degenerate with the $J/\psi$.  The BES
 has searched for this hypothetical particle in a $\rho\pi$ scan across
 the $J/\psi$ region in $e^{+}e^{-}$ annihilations as well as in decays
 $\psi(2S)\rightarrow\pi\pi O$, $O\rightarrow\rho\pi$,
 and found no evidence for its existence~\cite{besvp,xyh}.  The data 
 constrains the mass and width of the $O$ to the range
 $\mid m_{O}-m_{J/\psi}\mid < 80MeV$
 and $4MeV < \Gamma_{O} < 50MeV$~\cite{harris}.  This mass, as
 indicated in Ref.~\cite{cb}, is several hundred MeV lower than the lightest
 vector glueball observed in lattice simulations of QCD without dynamical
 quarks.  More recently, a few more experimental facts unfavorable to this
 model have been reported by BES.  One is the identification of 
 isospin-violating VP mode $\psi(2S)\rightarrow\omega\pi^{0}$
 with a large branching fraction~\cite{harris}.  This contradicts
 the essence of the model that the pattern of suppression is dependent
 on the spin-parity of the final state mesons.  The other is the finding
 of suppression of $\psi(2S)$ decays into vector plus tensor (VT)
 final states~\cite{besvt}.
 Since hadronic VT decays, unlike the VP decays, conserve HHC, some other
 mechanism must be responsible for this suppression in the model.
 Furthermore, it has been argued that the $O$ may also explain
 why $J/\psi$ decays to $\phi f_{0}$ (named previously $S^{\ast}$) but not to
 $\rho\delta$, since the $O$ mixes with the $\phi$ and enhances a
 mode that would be otherwise suppressed~\cite{blt}.  However, the
 observation of non-suppressed $\psi(2S)\rightarrow\phi f_{0}$~\cite{harris}, 
 which implies the absence of anomalous enhancement 
 in $J/\psi\rightarrow\phi f_{0}$, would rule out such an explanation.
 Anselmino et al. extended the idea of $J/\psi-O$
 mixing to the case of $\eta_c\rightarrow$VV and $p\overline{p}$~\cite{anselm}. 
 They suggested that the enhancement of these decays can be attributed to the
 presence of a trigluonium pseudoscalar state with a mass not far from the
 $\eta_c$ mass.  So far no experimental data has supported the existence of
 such a state.

 Recently Brodsky and Karliner proposed 
 the existence of intrinsic charm $\mid \overline{q}q\overline{c}c>$ 
 Fock components of the light vector mesons as another mechanism 
 to account for the $J/\psi$ decays to VP channels and their suppression 
 of $\psi(2S)$~\cite{bk}.  
 They also suggested comparing branching fractions for 
 the $\eta_c$ and $\eta_{c}(2S)$ as clues to the importance of $\eta_c$ intrinsic 
 charm excitations in the wavefunctions of light hadrons.  However, 
 the BES observation of $\psi(2S)\rightarrow\omega\pi^{0}$ would again 
 appear to disfavor this model.

 Chaichian and T\"{o}rnqvist suggested a model which invokes a form factor
 falling exponentially with the energy to suppress all $\psi(2S)$ decays
 to lowest-lying two-body meson final states~\cite{ct}.  However, the BES
 report on observation of a number of $\psi(2S)$ hadronic two-body decays
 such as $b_1 \pi$, $\phi f_0$, $K_{1}(1270)\overline{K}$,
 and $\omega\pi^{0}$ have proved the contrary~\cite{besap,harris}.
 In addition, the BES upper limit at 90\% C.L.
 $B(\psi(2S)\rightarrow\rho\pi) < 2.8\times 10^{-5}$~\cite{harris} 
 is well below the branching fraction predicted by this model, 
 $7\times 10^{-5}$.

 The generalized hindered M1 transition mechanism proposed by
 Pinsky~\cite{pnsky} relates the process
 $\psi(2S)\rightarrow\gamma\eta'$ to the hindered M1 transition
 $\psi(2S)\rightarrow\gamma\eta_c$.  This predicts
 $Q_{\gamma\eta'}$ to be $2\times 10^{-3}$, which, as already shown in
 Ref.~\cite{besrad},
 falls more than an order of magnitude below the BES data
 $(3.6\pm0.9)\times 10^{-2}$.
 According to this model, the hadronic decays of
 $\psi(2S)$ to VP final states
 are also a generalized hindered M1 transition.  The branching fraction
 for the decay of $\psi(2S)\rightarrow\rho\pi$ is estimated to be
 $4\times10^{-5}$, as compared to the measured limit of $2.8\times 10^{-5}$.  
 Moreover, it is inferred from this model that
 $\psi(2S)\rightarrow\gamma f_2$ decay should be suppressed whereas
 $\psi(2S)\rightarrow\omega f_2$ should not~\cite{chao}.  
 However, the experimental facts from BES contradict
 this assumption~\cite{besvt,harris}.

 Karl and Roberts have suggested a proposal to explain the $\rho\pi$
 puzzle based on the mechanism of sequential quark pair creation~\cite{kr}.
 Even though their predictions could generally accommodate the data for decays
 of $J/\psi$ and $\psi(2S)$ to $\rho\pi$ or to $K^{\ast}\overline{K}$, 
 it seems hard to  explain the large branching fraction for $\phi$ decays to 
 $\rho\pi$~\cite{pdg} 
 due to the fact that their fragmentation probability tends to zero
 as the mass of the  $\rho\pi$ decaying system approaches 1GeV.

 More recently, Li et al.~\cite{lbz} pointed out that final-state interactions
 in $J/\psi$ and $\psi(2S)$ decays give rise to effects which are of the
 same order as the tree level amplitudes, and may be a possible explanation
 for all the observed suppressed modes of $\psi(2S)$ decays including
 $\rho\pi$, $K^{\ast}\overline{K}$ and $\omega f_2$.
 They thus predicted qualitatively large production rates of $a_{1}\rho$ and
 $K_{1}^{\ast}\overline{K}^{\ast}$ for $\psi(2S)$, the verification of which may
 give further support to their model.  So far, BES has never reported such
 measurements; nevertheless, useful information
 on $a_{1}\rho$ and $K_{1}^{\ast}\overline{K}^{\ast}$ could be obtained
 from its published data as shown in Ref.~\cite{besvt}.
 The lack of evidence within the invariant mass distribution plots
 (see Fig.3 and Fig.5 of Ref.~\cite{besvt}) that the $\rho\pi$ recoiled against a $\rho$ 
 for events of  $\psi(2S)\rightarrow\rho^{0}\rho^{\pm}\pi^{\mp}$ 
 and that $\pi^{\pm}K^{\mp}$ recoiled against a $K^{\ast 0}$ for events of 
 $\psi(2S)\rightarrow\pi^{+}\pi^{-}K^{+}K^{-}$ suggests that they are unlikely to be the favored 
 modes in $\psi(2S)$ decays.

 A model put forward by G\'{e}rald and Weyers entertains the assumption that
 the three-gluon annihilation amplitude and the QED amplitude add incoherently
 in all channels in $J/\psi$ decays into light hadrons, while in the case of
 $\psi(2S)$ decays the dominant QCD annihilation amplitude is not into three
 gluons, but, via a two step process, into a specific configuration of five
 gluons~\cite{gw}.
 Besides explaining the measurements on $\psi(2S)$ decays to $\rho\pi$,
 $K^{\ast}\overline{K}$, and $\omega\pi^{0}$, 
 this model predicts a sizeable 
 $\psi(2S)\rightarrow$ ($\pi^{+}\pi^{-}$ or $\eta$)$h_{1}(1170)$
 branching fraction.  Indeed the BES has performed
 extensive analysis of decays $\psi(2S)\rightarrow\pi\pi\rho\pi$ 
 to look for new particles; however,
 it is unlikely that a conclusive signal for $h_{1}(1170)$ has ever been 
 observed in the inclusive spectrum of $\psi(2S)$ decays 
 to $\pi\pi\rho\pi$~\cite{xyh}.

 Chen and Braaten proposed an explanation~\cite{cb} for
 the $\rho\pi$ puzzle, arguing that the decay $J/\psi\rightarrow\rho\pi$
 is dominated by a Fock state in which the $c\overline{c}$ pair is
 in a color-octet $^{3}S_{1}$ state which decays via
 $c\overline{c}\rightarrow q\overline{q}$, while the suppression of
 this decay mode for the $\psi(2S)$ is attributed to a
 dynamical effect due to the small energy gap between the mass
 of the $\psi(2S)$ and the $D\overline{D}$ threshold. Using the BES data
 on the branching fractions into $\rho\pi$ and $K^{\ast}\overline{K}$ as
 input, they predicted the branching fractions for many other
 VP decay modes of the $\psi(2S)$.  Most recently, Feldmann and Kroll
 parametrized the strong interaction mechanism for the hadron-helicity
 non-conserving decays in a similar way, but interpreted it
 differently~\cite{fk}.  They argued that, for these processes,
 the charmonium state decays through a light-quark Fock component
 by a soft mechanism, which is characteristic of OZI-rule allowed
 strong decays.  Estimating the light-quark admixture by meson-mixing,
 they also obtained a numerical description of the branching fractions
 for many VP decay modes of the $J/\psi$ and $\psi(2S)$.
 The predictions of both models are in good agreement with the measured
 branching fractions (some are preliminary) from the BES 
 experiment~\cite{harris} as well as the PDG data~\cite{pdg}.
 Chen and Braaten's proposal also has implications for the
 angular distributions for two-body decay modes of $\psi(2S)$.
 Nevertheless, such measurements would be extremely difficult, if not 
 impossible, to perform for those strongly suppressed decay modes in 
 $\psi(2S)$ decays.  Feldmann and Kroll, on the other hand, has extended 
 their mixing approach to the $\eta_c\rightarrow$ VV decays and obtained 
 a reasonable description of the branching fractions for these decays 
 while the $\eta_{c}(2S)\rightarrow$ VV decays are expected 
 to be strongly suppressed. 

 From the above discussion we see that essentially none of the models are
 able to explain all known experimental results; in particular no analysis
 on the suppression of $\psi(2S)\rightarrow$ VT decays has been given.
 Not a few models appear to have more assumptions than predictions, not to
 mention quantitative predictions.  While the current data seem to rule out 
 convincingly some of the models, a few other models may warrant further 
 consideration; for them both detailed theoretical analyses and additional
 experimental tests are demanded.  It seems to us that a key premise for
 physical considerations is to establish
 whether the $J/\psi$ decays or the $\psi(2S)$ decays are anomalous.
 An amplitude analysis made for the two-body decays of $J/\psi$ 
 to VP~\cite{suz} has shown that nothing anomalous is found in the magnitudes 
 of the three-gluon and one-photon decay amplitudes.  If this is sustained, 
 those arguments presupposing the $J/\psi$ as the origin of the anomaly 
 should be disregarded. 

 In summary, we have examined the $\rho\pi$ puzzle of $J/\psi$ and
 $\psi(2S)$ decays in the light of current experimental data.  The estimates
 of the ratio of $\psi(2S)$ to $J/\psi$ hadronic decay rates, using two different
 approaches, differ substantially from each other.  The one using only 
 the data of exclusive hadronic decays appears to be underestimated.
 The other estimate, that is based on computation of the inclusive ggg decay
 rate by subtracting other decay rates from the total decay rate, differs 
 by 2$\sigma$-error of this estimate from the naive prediction of PQCD,
 even though its central value is about a factor of two as large as the latter.
 By comparing this estimated ratio, and taking account the corrections
 associated with specific exclusive decay modes with the corresponding
 experimental data, anomalies in $J/\psi$ and $\psi(2S)$ decays to VP, VT,
 AP and some other final states are evident.  We found from our analysis
 that the exclusive hadronic decays of the $\psi(2S)$ so far reconstructed
 experimentally account for only a small fraction of the total $\psi(2S)$
 decays and a ratio of $\psi(2S)$ to $J/\psi$ hadronic decay rates that 
 exceeds the estimated value is expect to occur for a considerable part of 
 the remaining $\psi(2S)$ decay channels. 
 We have also shown that the recent new results from the BES experiment 
 provide crucial tests of various theoretical models proposed to understand 
 the $\rho\pi$ puzzle.  Further experimental and theoretical efforts 
 are required, in order to fill the missing data and definitively solve the
 perplexing $\rho \pi$ puzzle.

 \vspace{0.25in}

The authors  wish to thank ~M.~Wise, ~F.~Porter, ~S.~J.~Brodsky, 
~G.~D.~Zhao, ~S.~F.~Tuan and ~J.~Weyers for enlightening discussions. 
The discussions with the BES collaboration are appreciatively acknowledged. 
This work was supported by the National Natural Science Foundation of 
China under Contract No. 19290400, the Chinese Academy of Sciences under
Contracts No. H-10 and No. E-01,  and by the U.S. Department of Energy 
under Contract No. DE-FG03-86ER40271.


\vskip 10cm

 Table 1.  Experimental data on branching fractions for electromagnetic 
 decays of $J/\psi$ and $\psi(2S)$ used in our analysis.
 All data are taken from PDG~\cite{pdg} except the branching fraction for
 $\psi(2S)\rightarrow\tau^+\tau^-$ which is a first measurement 
 by BES~\cite{tau}.
\vskip 1cm
\begin{center}
\begin{tabular}{|c|c|c|}\hline
Channel & B($J/\psi$) & B($\psi(2S)$) \\\hline
$\gamma^*\rightarrow hadrons$ & $(17.0\pm2.0)\%$ & $(2.9\pm0.4)\%$ \\\hline
$e^+e^-$ & $(5.93\pm0.10)\%$ & $(8.8\pm1.3)\times 10^{-3}$ \\\hline
$\mu^+\mu^-$ & $(5.88\pm0.10)\%$ & $(1.03\pm0.35)\%$ \\\hline
$\tau^+\tau^-$ &               & $(2.71\pm0.70)\times 10^{-3}$ \\\hline
\end{tabular}
\end{center}
\vskip 2cm


 Table 2.  Experimental data on branching fractions for 
 $J/\psi$ and $\psi(2S)$
 decays to lower mass charmonium states used in our analysis.        
 All data are taken from PDG~\cite{pdg}. 

\begin{center}
\begin{tabular}{|c|c|c|}\hline
Channel & B($J/\psi$) & B($\psi(2S)$) \\\hline
$\gamma\eta_{c}$ & $(1.3\pm0.4)\%$ & $(0.28\pm0.06)\%$  \\\hline
$\pi^+\pi^-J/\psi$ &         & $(31.0\pm2.8)\%$ \\\hline
$\pi^0\pi^0J/\psi$ &         & $(18.2\pm2.3)\%$ \\\hline
$\eta J/\psi$ &          & $(2.7\pm0.4)\%$ \\\hline
$\pi^0J/\psi$ &          & $(9.7\pm2.1)\times 10^{-4}$ \\\hline
$\gamma\chi_{c0}$ &          & $(9.3\pm0.9)\%$ \\\hline
$\gamma\chi_{c1}$ &          & $(8.7\pm0.8)\%$ \\\hline
$\gamma\chi_{c2}$ &          & $(7.8\pm0.8)\%$ \\\hline
\end{tabular}
\end{center}
\vskip 2cm

 Table 3.  Branching fractions for the $J/\psi$ and $\psi(2S)$ exclusive 
 hadronic decays used in our analysis. 
 All data are from PDG~\cite{pdg}. 

\begin{center}
\begin{tabular}{|c|c|c|}\hline
Mode & B($J/\psi$) & B($\psi(2S)$) \\\hline
$\pi^+\pi^-\pi^0$ & $(1.50\pm0.20)\%$ & $(8\pm5)\times 10^{-5}$ \\\hline
$2(\pi^+\pi^-)\pi^0$ & $(3.37\pm0.26)\%$ & $(3.0\pm0.8)\times 10^{-3}$ \\\hline
$3(\pi^+\pi^-)\pi^0$ & $(2.9\pm0.6)\%$ & $(3.5\pm1.6)\times 10^{-3}$ \\\hline
$K^+K^-\pi^+\pi^-$ & $(7.2\pm2.3)\times 10^{-3}$ & $(1.6\pm0.4)\times 10^{-3}$ \\\hline
$p\overline{p}\pi^+\pi^-$ & $(6.0\pm0.5)\times 10^{-3}$ &
$(8.0\pm2.0)\times 10^{-4}$ \\\hline
$p\overline{p}\pi^0$ & $(1.09\pm0.09)\times 10^{-3}$ & $(1.4\pm0.5)\times 10^{-4}$ \\\hline
$p\overline{p}$ & $(2.12\pm0.10)\times 10^{-3}$ & $(1.9\pm0.5)\times 10^{-4}$ \\\hline
$K^+K^-$ & $(2.37\pm0.31)\times 10^{-4}$ & $(1.0\pm0.7)\times 10^{-4}$ \\\hline  
\end{tabular}
\end{center}
\vskip 4cm

\newpage      

Table 4.  The ratio of branching fractions of $\psi(2S)$ and $J/\psi$ 
 exclusive decays: $Q_f = B(\psi(2S))/B(J/\psi)$.
 All data are from PDG~\cite{pdg}. 
 Upper limits are given at the $90\%$ confidence level.
\vskip 1cm

\begin{tabular}{|c|c|c|c|c|c|}\hline
mode &  HHC &  orbital momentum  & 
 pred. $Q_f$ $(\%)$ & meas. $Q_f$ $(\%)$ \\\hline 

$P\overline{P}$  & 1 & 1 & $24.0\pm 5.6$ &  $9.0\pm2.4$ \\\hline 

$\rho\pi$  & 0.71 & 1  & $17.0\pm4.0$ & $<0.65$ \\\hline

$K^{+}\overline{K}^{*}(892)^{-}$ &  0.71 & 1  & $17.0\pm4.0$ & $<1.1$ \\\hline

$\omega f_2(1270)$  & 1 & 0.71  & $17.0\pm4.0$ & $<4.0$ \\\hline

$\rho a_2(1320)$  & 1 & 0.71 & $17.0\pm4.0$ & $<2.1$ \\\hline

$K^{\ast}(892)^{0}\overline{K}_2^{\ast}(1430)^0$  & 
1 & 0.71  & $17.0\pm4.0$ & $<1.8$ \\\hline 

$\phi f_2'(1525)$  & 1 & 0.71 & $17.0\pm4.0$ & $<5.6$ \\\hline 

$b_1^{\pm}\pi^{\mp}$  & 1 & 0.71& $17.0\pm4.0$ & $17.3\pm5.2$ \\\hline

$K_1^{\pm}(1270) K^{\mp}$  & 1 & 0.71& $17.0\pm4.0$ & $>33.3$  \\\hline

$K_1^{\pm}(1400) K^{\mp}$  & 1 & 0.71& $17.0\pm4.0$ & $<8.2$  \\\hline

$\gamma\eta'(958)$  & 1  & 1 & $24.0\pm5.6$ & $3.5\pm1.0$  \\\hline

$\gamma\eta$  & 1 & 1  & $24.0\pm5.6$ & $<10.5$ \\\hline
\end{tabular}


\begin{thebibliography}{99}
\bibitem{com} More accurate values for the branching
 fractions of $J/\psi, \psi(2S)\rightarrow \gamma^*\rightarrow hadrons$ 
 can be obtained by making the perturbative correction of
 $(1+\overline{\alpha}_{s}/\pi)$ to the leptonic branching fractions.  
 However, we do not use them to replace the old measurements quoted in PDG~\cite{pdg} 
 due to the poor knowledge of the theoretical uncertainties.
\end{thebibliography}
\end{document}